\documentclass[english,reprint, aps, prd, showkeys , showpacs]{revtex4-1}
\usepackage[T1]{fontenc}
\usepackage[utf8x]{inputenc}
\usepackage{prettyref}
\usepackage{calc}
\usepackage{amsmath}
\usepackage{amssymb}
\usepackage{hyperref}

\makeatletter
\usepackage{subfigure}
\usepackage{prettyref}

\newrefformat{tab}{Table\,\ref{#1}}
\newrefformat{fig}{Figure\,\ref{#1}}
\newrefformat{sub}{Section\,\ref{#1}}
\newrefformat{subsec}{Section\,\ref{#1}}
\newrefformat{app}{Appendix\,\ref{#1}}
\newrefformat{subfig}{Figure\,\ref{#1}}
\usepackage[capitalize]{cleveref}
\newrefformat{eq}{Eq.\,\textup{(\ref{#1})}}

\makeatother

\newcommand{\hcs}{h.c.^{\!\!*}}

\usepackage{babel}
\begin{document}

\title{Gauge dependence of tadpole and mass renormalization for a seesaw
extended 2HDM}

\author{Vytautas Dūdėnas}
\email{vytautasdudenas@inbox.lt}

\author{Thomas Gajdosik}

\affiliation{Institute of theoretical Physics and Astronomy, Faculty of Physics,
Vilnius University.}

\begin{abstract}
We study the gauge dependence of the neutrino mass renormalization
in a two Higgs doublet model, that is extended with one singlet seesaw
neutrino. This model gives only one light neutrino a mass at tree
level, while the second light mass is generated at loop level via
the interaction with the second Higgs doublet. At one loop level,
one neutrino stays massless. We use multiplicative renormalization
constants to define counterterms. The renormalized mass parameters
are defined as the complex poles of the propagators, using the complex
mass scheme for mass renormalization. With this setup, we analytically
get the expressions for the neutrino mass counterterms and isolate
the gauge dependent part. We show, how relating this gauge dependent
part with the tadpole renormalization leads to gauge independent counterterm
definitions, hence gauge independent bare masses for neutrinos. 
\end{abstract}


\maketitle

\section{Introduction}
Neutrino oscillations are known for more than 30 years \citep{Heeger:2004mp}.
They prove that neutrinos are not massless. However, how exactly neutrinos
get their masses in the framework of quantum field theory is still
unclear. Seesaw mechanisms \citep{Molinaro:2013toa,Orloff:2005nu}
are by far the most popular attempts to extend the Standard model
with massive neutrinos. The type I seesaw mechanism \citep{Molinaro:2013toa}
is the earliest and simplest such extension, which includes neutrino
mass terms induced by the Higgs boson of the Standard model (SM).
In case there are more Higgs bosons than the single SM Higgs, the
type I seesaw extension can be generalized as in \citep{Grimus:1989pu}.
This allows for a wider range of configurations in the seesaw and
Yukawa sectors to generate the masses for neutrinos that are in agreement
with the experimental values. Also, there are numerous theoretical
motivations \citep{Haber:1984rc,Kim:1986ax,Lee:1973iz,Peccei:1977hh}
suggesting a larger scalar sector. We restrict ourselves to a general
CP conserving two Higgs doublet model (2HDM) \citep{Branco:2011iw},
which can be viewed as a general class of more specific models that
include two scalar doublets under the gauge group $SU\left(2\right)_{\text{weak}}$.

The 2HDM paired with the seesaw mechanism gives a new way of generating
masses for neutrinos that is absent in the usual SM seesaw extensions.
That is, the mass terms that are absent at tree level arise at loop
level due to the interactions with the second Higgs doublet. This
radiative mass generation makes it possible to account for both experimentally
measured mass differences at one loop level having only one sterile
neutrino in the seesaw mechanism. This set up, with the 2HDM and one
sterile neutrino at one loop was first proposed in \citep{Grimus:1989pu}
and we call it the Grimus-Neufeld model (GN model). 

We look at the gauge parameter dependence of the neutrino mass renormalization
in this GN model with a CP symmetric 2HDM potential. It is proven
in general \citep{Gambino:1999ai}, that the position of the complex
pole of the propagator is independent of the gauge. Hence one can
extend the on-shell (OS) scheme to the complex domain to define gauge
invariant masses as is done in the complex mass scheme (CMS) \citep{Denner:1999gp,Denner:2006ic}.
However, this doesn't mean that the mass counterterms are necessarily
gauge parameter independent. In fact, at one loop there is the same
gauge dependence of the mass counterterms in the CMS as in the OS scheme.
This is because the one loop expressions for the OS are the same as
in the CMS except for the required reality of loop functions in the
OS scheme. As long as the mass is evaluated at the exact pole (as
in the CMS), this gauge dependence of the counterterm doesn't bother
the definition of mass since the exact pole is gauge independent anyway.
Defining a gauge independent counterterm, however, is important in
other schemes such as (modified) minimal subtraction, where the gauge
dependence might occur in the running of parameters \citep{Denner:2016etu, Altenkamp:2017ldc}.
Hence it is worth to look at the possibilities to define gauge independent
mass counterterms in the CMS or the OS, as well. 

In the GN model, we analytically check that the gauge dependent terms
for the fermion two point function vanish if the tadpole diagrams
are attached to the propagator as discussed in \citep{Liebler:2010bi}.
This way of dealing with gauge dependent parts originates from the
pinch technique \citep{Cornwall:2010upa}. Hence applying this technique
to define numerically gauge invariant counterterms seems rather straightforward.
However, to analytically isolate these tadpole diagrams from the counterterms
requires some effort. We present how we achieve this isolation of
the gauge dependent terms for the neutrino mass counterterms in the
GN model. We try to be as transparent as possible in showing our steps
so that the reader can easily reproduce our results. All our renormalization
constants arise from multiplicative renormalization and we use Weyl
spinors for our expressions rather than Dirac spinors. 

In \prettyref{sec:Definitions-and-the} we present the main definitions
and discuss the implications of using the complex mass scheme over
the on-shell scheme. In \prettyref{sec:Scalar-sector-and} we introduce
the scalar sector and present the tadpole renormalization conditions
in the 2HDM. In \prettyref{sec:Yukawa-sector} we introduce the Yukawa
sector of the GN model and show the expressions of mass counterterms
for neutrinos. The relationship between tadpole conditions of \prettyref{sec:Scalar-sector-and}
and mass counterterms is also explained in \prettyref{sec:Yukawa-sector}.
In \prettyref{sec:Ariving-at-the} we show how we set up the calculations
using Sarah\citep{Staub:2008uz}, FeynArts\citep{Hahn:2000kx} and
FormCalc\citep{Hahn:1998yk} and present the analytical results. \prettyref{sec:Ariving-at-the}
is accompanied by the \prettyref{app:appendix1} in which we present
some intermediate steps of the derivations. We conclude the results
in \prettyref{sec:Gauge-dependence-of-1} by discussing the cancellation
of the gauge dependence of neutrino propagators in the GN model. 

\section{Definitions and the complex mass scheme\label{sec:Definitions-and-the}}

We use the same definitions as in \citep{Dudenas:2017}, where we
presented the adaptation of the complex mass scheme \citep{Denner:2006ic}
for Majorana fermions in Weyl spinor formalism. The renormalized Green
functions are:
\begin{align}
\left\langle \phi_{1}...\phi_{n}\right\rangle _{1PI}^{\left[loop\right]} & =\frac{\delta^{n}\hat{\Gamma}^{\left[loop\right]}}{\delta\phi_{1}..\delta\phi_{n}}\Big|_{\phi_{i}=0}\equiv\hat{\Gamma}_{\phi_{1}...\phi_{n}}^{\left[loop\right]}\nonumber \\
 & \equiv\Gamma_{\phi_{1}...\phi_{n}}^{\left[loop\right]}+\delta\Gamma_{\phi_{1}...\phi_{n}}^{\left[loop\right]}\,,\label{eq:greens}
\end{align}
where $\delta\Gamma^{\left[loop\right]}$ stands for the counterterm
part of the renormalized effective action. The superscript denotes
the loop order of the function in consideration. The tadpole function
is defined as the special case of \prettyref{eq:greens}:
\begin{equation}
T_{\phi}^{\left[loop\right]}\equiv\Gamma_{\phi}^{\left[loop\right]}\,.\label{eq:tadpole definition}
\end{equation}
The definitions for using Weyl spinors as the basis of Feynman diagram
calculations can be found in \citep{Dreiner:2008tw}. The scalar parts
of Green's functions of a left handed Weyl spinor $\nu_{i}$ and its
hermitian conjugate $\nu_{i}^{\dagger}$ can be separated by the Lorentz
index structure: 
\begin{align}
 & \hat{\Gamma}_{\nu_{i}\nu_{i}}=m_{i}\hat{\Sigma}_{\nu_{i}\nu_{i}}\,,\,\,\,\hat{\Gamma}_{\nu_{i}^{\dagger}\nu_{i}^{\dagger}}=m_{i}\hat{\Sigma}_{\nu_{i}^{\dagger}\nu_{i}^{\dagger}}\,,\nonumber \\
 & \hat{\Gamma}_{\nu_{i}\nu_{j}^{\dagger}}=p\sigma\hat{\Sigma}_{\nu_{i}\nu_{j}^{\dagger}}\,,\,\,\,\hat{\Gamma}_{\nu_{i}^{\dagger}\nu_{j}}=p\bar{\sigma}\hat{\Sigma}_{\nu_{i}^{\dagger}\nu_{j}}\,.\label{eq:selfs}
\end{align}
The definitions of \prettyref{eq:selfs} work well for the on-shell
scheme, but have to be slightly modified for the complex mass scheme.

We work in renormalized perturbation theory, where the renormalized
parameters $p$ and the renormalized fields $\phi_{j}$ are related
to bare parameters and bare fields by multiplicative renormalization
constants:
\begin{equation}
p_{0}=p\left(1+\delta_{p}\right)\,,\,\,\phi_{0i}=\sum_{j}\left(1_{ij}+\delta_{ij}\right)\phi_{j}\,.\label{eq:rescales}
\end{equation}
We use the subscript $0$ to denote the bare quantities, $1_{ij}$
stands for the Kronecker delta, $\delta_{p}$ and $\delta_{ij}$ are
one loop order renormalization constants. These redefinitions of parameters
and fields give rise to the counterterms $\delta\Gamma_{\phi_{1}...\phi_{n}}^{\left[loop\right]}$
in \prettyref{eq:greens}. 

We use the general $R_{\xi}$ gauge for calculations. As we will look
at the gauge parameter dependencies, we will frequently look at only
the gauge parameter dependent part of the expressions. To denote the
gauge dependent term, we will add the gauge parameter $\xi$ in the
subscript at the end of the renormalization constants, self energies
and tadpole functions; for example:
\begin{equation}
\delta_{p}\equiv\delta_{p\xi}+\mbox{gauge independent terms},\,\,\delta_{p\xi}=\delta_{p\xi_{W}}+\delta_{p\xi_{Z}}\,.\label{eq:Gauge dependence notation}
\end{equation}

We use the complex mass scheme \citep{Denner:2006ic} (CMS) to renormalize
masses and fields. The CMS for mixed fermions is presented in \citep{Kniehl:2008cj,Kniehl:2014gfa,Espriu:2002xv}
and the adaptation to Weyl spinor formulation is presented in \citep{Dudenas:2017}.
Here we mention the main differences that need to be considered when
generalizing the OS framework to the CMS. Considering a Majorana mass
term for the Weyl fermion $\nu$: 
\begin{equation}
\mathcal{L}_{m_{0}}=-\frac{1}{2}m_{0}\nu_{0}\nu_{0}-\frac{1}{2}m_{0}^{\dagger}\nu_{0}^{\dagger}\nu_{0}^{\dagger}\,,\label{eq:majorana mass term}
\end{equation}
the Majorana phase can be adjusted, so that $m_{0}\in\mathbb{R}$:
\begin{equation}
\mathcal{L}_{m_{0}}=-\frac{1}{2}m_{0}\left(\nu_{0}\nu_{0}+\nu_{0}^{\dagger}\nu_{0}^{\dagger}\right)\,.\label{eq:majorana mass term real}
\end{equation}
Renormalizing the mass parameter leads to
\begin{equation}
\mathcal{L}_{m_{0}}=-\frac{1}{2}m\left(\nu_{0}\nu_{0}+\nu_{0}^{\dagger}\nu_{0}^{\dagger}\right)+c.t.\,,
\end{equation}
where $m\in\mathbb{C}$ and $c.t.$ stands for the counterterms. Hence
the CMS introduces an apparent non hermitcity in the renormalized
tree level Lagrangian (the full Lagrangian including all the counterterms
is hermitian). Also, the condition for the residue at the complex
pole leads to an additional phase difference in the fields \citep{Kniehl:2008cj,Kniehl:2014gfa,Espriu:2002xv}.
That means that the field renormalization constants are not hermitian
conjugate to each other either \citep{Dudenas:2017}: 
\begin{equation}
\nu_{0}^{\dagger}=\left(1+\bar{\delta}\right)\bar{\nu}\,,\,\,\nu_{0}=\left(1+\delta\right)\nu\,\Rightarrow\,\bar{\nu}\neq\nu^{\dagger}\,,\,\,\delta^{\dagger}\neq\bar{\delta}\,,\label{eq:field renormalization}
\end{equation}
where we use overbars as parts of the names of the renormalization
constants and the fields. Hence the renormalized mass Lagrangian in
the CMS is: 
\begin{equation}
\mathcal{L}_{m}=-\frac{1}{2}m\left(\nu\nu+\bar{\nu}\bar{\nu}\right)\,.\label{eq:renormalized mass lagrangian}
\end{equation}
Comparing with the bare Lagrangian, we see that we could write \prettyref{eq:majorana mass term}
or \prettyref{eq:majorana mass term real} as: 
\begin{equation}
\mathcal{L}_{m_{0}}=-\frac{1}{2}m_{0}\nu_{0}\nu_{0}+h.c.
\end{equation}
We cannot write \prettyref{eq:renormalized mass lagrangian} in the
same way, since it is not hermitian. However, we can try to define
a new symbol $\hcs$ to have the possibility to write:
\begin{equation}
\mathcal{L}_{m}=-\frac{1}{2}m\left(\nu\nu+\bar{\nu}\bar{\nu}\right)=-\frac{1}{2}m\nu\nu+\hcs \label{eq:renormalized mass lagrangian-1}
\end{equation}
In this equation the symbol $\hcs$ makes the replacement for
the field $\nu\to\bar{\nu}$ and leaves $m\to m$. The mass parameter
is unchanged in the $\hcs$ since we found the basis, in which
the bare parameter is real by absorbing the phase into $\nu_{0}$
in \prettyref{eq:majorana mass term real}. Hence the algebraic structure
of \prettyref{eq:majorana mass term real} is kept in the renormalized
version shown in \prettyref{eq:renormalized mass lagrangian}. A similar
thing happens in the CP conserving Higgs sector: the CP symmetry constrains
the form of the Lagrangian, which has to be kept during the renormalization
condition. Also, in the scalar and the vector case, if we have $\phi_{0}\in\mathbb{R}$,
then $\phi=\bar{\phi}$. The easiest way to generalize the $\hcs$
symbol is to say that we choose the basis in which the bare parameters
that can be real are made real; then we can summarize: 
\begin{equation}
\hcs:\,\begin{cases}
p\to p\,,\,\,\phi\to\bar{\phi}\,; & p_{0}\in\mathbb{R}\\
p\to p^{\dagger}\,,\,\phi\to\bar{\phi}\,; & p_{0}\not\in\mathbb{R}\;.
\end{cases}\label{eq:hc star}
\end{equation}
Normally, if a bare parameter is related to the bare mass term, that
parameter can be made real by absorbing the phase into the field.
Hence the second line of \prettyref{eq:hc star} assumes that there
is no effect of the mass renormalization to the parameter $p$ if
$p_{0}$ cannot be related to the mass term. While this assumption
is correct at one loop level, the definition \prettyref{eq:hc star}
at higher loops should be treated with caution. Without going into
too much technical details, one can think of $\hcs$ as a shorthand
notation for the renormalized $h.c.$ terms of the bare Lagrangian. 

Now we can come back to the definitions of \prettyref{eq:selfs}.
As the CMS renormalized field is $\bar{\nu}$ and not $\nu^{\dagger}$,
as can be seen from \prettyref{eq:field renormalization}, we write
\citep{Dudenas:2017}: 
\begin{align}
 & \hat{\Gamma}_{\nu_{i}\nu_{i}}=m_{i}\hat{\Sigma}_{\nu_{i}\nu_{i}}\,,\,\,\,\hat{\Gamma}_{\bar{\nu}_{i}\bar{\nu}_{i}}=m_{i}\hat{\Sigma}_{\bar{\nu}_{i}\bar{\nu}_{i}}\,,\nonumber \\
 & \hat{\Gamma}_{\nu_{i}\bar{\nu}_{j}}=\,p\sigma\hat{\Sigma}_{\nu_{i}\bar{\nu}_{j}}\,,\,\,\,\hat{\Gamma}_{\bar{\nu}_{i}\nu_{j}}=p\bar{\sigma}\hat{\Sigma}_{\bar{\nu}_{i}\nu_{j}}\,.\label{eq:selfs-1}
\end{align}
The difference between \prettyref{eq:selfs} and \prettyref{eq:selfs-1}
is rather formal: i.e. one doesn't really see the difference when
calculating the Feynman diagrams. However, for using the CMS for field
and mass renormalization, one should keep this difference in mind
for the conceptual consistency. 

After we have the consistent set up for renormalizing the fermions
in the CMS, we continue to look at the gauge parameter dependencies
of the renormalization constants in this scheme. The multiplicative
renormalization constants \prettyref{eq:rescales} can be used for
any renormalization condition. The algebra of the CMS is basically
the same as in the OS, as the CMS is just the analytical continuation
of the OS to the complex domain. In this paper, we study the algebraic
relations that allow to isolate the gauge parameter term in the mass
counterterm. As this procedure is purely algebraic, the expressions
concerning the isolation of the gauge dependent part are the same
as in the OS scheme apart from the reality requirement. We, however,
do these manipulations with the CMS in mind, as the generalizations
despite being rather straightforward are still needed for a full consistency.
We now turn to the explicit expressions for the GN model. 

\section{Scalar sector and tadpole conditions\label{sec:Scalar-sector-and}}

The general 2HDM is an extension of the SM with a second Higgs doublet
having the same charges as the SM Higgs doublet. The most general
potential can be written as \citep{Branco:2011iw,Davidson:2005cw}:
\begin{widetext}
\begin{align}
\mathcal{V}_{Higgs} & =m_{011}^{2}H_{01}^{\dagger}H_{01}+m_{022}^{2}H_{02}^{\dagger}H_{02}-(m_{012}^{2}H_{01}^{\dagger}H_{02}+h.c.)\nonumber \\
 & +\frac{1}{2}\lambda_{01}(H_{01}^{\dagger}H_{01})^{2}+\frac{1}{2}\lambda_{02}(H_{02}^{\dagger}H_{02})^{2}+\lambda_{03}(H_{01}^{\dagger}H_{01})(H_{02}^{\dagger}H_{02})+\lambda_{04}(H_{02}^{\dagger}H_{01})(H_{01}^{\dagger}H_{02})\nonumber \\
 & +\Big[\frac{1}{2}\lambda_{05}(H_{02}^{\dagger}H_{01})(H_{02}^{\dagger}H_{01})+\lambda_{06}(H_{01}^{\dagger}H_{01})(H_{01}^{\dagger}H_{02})+\lambda_{07}(H_{02}^{\dagger}H_{02})(H_{02}^{\dagger}H_{01})+h.c.\Big]\,,\label{eq:L higgs bare}
\end{align}
\end{widetext}
where $H_{01}$ and $H_{02}$ are the two Higgs doublets. In a general
basis, they both develop VEVs: $v_{01}$ and $v_{02}$, respectively.
The VEV value that is responsible for the electroweak symmetry breaking
is $v_{0}^{2}=v_{01}^{2}+v_{02}^{2}$. We choose to work in the Higgs
basis, where we can parametrize the Higgs doublets as:
\begin{align}
H_{01} & =\left(\begin{array}{c}
\chi_{0W}^{+}\\
\frac{1}{\sqrt{2}}\left(v_{0}+h_{0}+i\chi_{0Z}\right)
\end{array}\right)\,,\nonumber \\
H_{02} & =\left(\begin{array}{c}
H_{0}^{+}\\
\frac{1}{\sqrt{2}}\left(H_{0}+iA_{0}\right)
\end{array}\right)\,.\label{eq:Higgs basis}
\end{align}
In this basis, $H_{02}$ is chosen to have $0$ vacuum expectation
value (VEV), $v_{0}$ is the VEV of $H_{01}$, $\chi_{0Z}$ and $\chi_{0W}$
stand for Goldstone bosons, $h_{0}$, $H_{0}$ and $A_{0}$ are neutral
scalars and $H_{0}^{+}$ is a charged scalar. Note that when we choose
the Higgs basis by \prettyref{eq:Higgs basis} and insert into the
\prettyref{eq:L higgs bare}, the parameters in \prettyref{eq:L higgs bare}
are the Higgs basis parameters and not the ones of the general basis.
The transformation of parameters between the Higgs and the general
basis can be found in \citep{Branco:2011iw,  Haber:2006ue}. We consider the CP conserving
case, where all the bare parameters are real,
\begin{equation}
m_{0ij}^{2},\lambda_{0k}\in\mathbb{R}\,;\,\,i,j=1,2\,,\,k=1,...,7\,,\label{eq:CP conserving}
\end{equation}
by an imposed CP symmetry on the bare Lagrangian.

After introducing the renormalization constants, \prettyref{eq:rescales},
we write the zeroth order renormalized effective action (or the renormalized
Lagrangian, ignoring the kinetic terms) of the Higgs sector as:
\begin{widetext}
\begin{align}
\Gamma_{\,\,\text{Higgs}}^{\left[0\right]} & =-m_{11}^{2}\bar{H}_{1}H_{1}-m_{22}^{2}\bar{H}_{2}H_{2}+\big\{ m_{12}^{2}\bar{H}_{1}H_{2}+\hcs\big\}\nonumber \\
 & -\frac{1}{2}\lambda_{1}(\bar{H}_{1}H_{1})^{2}-\frac{1}{2}\lambda_{2}(\bar{H}_{2}H_{2})^{2}-\lambda_{3}(\bar{H}_{1}H_{1})(\bar{H}_{2}H_{2})-\lambda_{4}(\bar{H}_{2}H_{1})(\bar{H}_{1}H_{2})\nonumber \\
 & -\Big[\frac{1}{2}\lambda_{5}(\bar{H}_{2}H_{1})(\bar{H}_{2}H_{1})+\lambda_{6}(\bar{H}_{1}H_{1})(\bar{H}_{1}H_{2})+\lambda_{7}(\bar{H}_{2}H_{2})(\bar{H}_{2}H_{1})+\hcs\Big]\,,\label{eq:L hermitian}
\end{align}
\end{widetext}
where we used the definitions of \prettyref{eq:hc star}. As the bare
fields $h_{0}$, $H_{0}$, $A_{0}$ are real, the renormalized fields are
written as:
\begin{align}
H_{1} & =\left(\begin{array}{c}
\chi_{W}^{+}\\
\frac{1}{\sqrt{2}}\left(v+h+i\chi_{Z}\right)
\end{array}\right)\,,\,\,H_{2}=\left(\begin{array}{c}
H^{+}\\
\frac{1}{\sqrt{2}}\left(H+iA\right)
\end{array}\right)\,,\nonumber \\
\bar{H}_{i} & =H_{i}^{T}\left(+\to-,\,i\to-i\right)\,.\label{eq:Higgs basis-1}
\end{align}
$\chi_{W}^{+}$ and $\chi_{W}^{-}$ are related to $\chi_{0W}^{+}$
as described by \prettyref{eq:field renormalization}. The same
holds for $H^{+}$ and $H^{-}$. The neutral fields appear in the
barred doublets in the same way as in the unbarred doublets. 

To get the minimum of the potential, \prettyref{eq:L hermitian}, we
need to solve three tadpole equations for the three neutral scalars.
It is important to note that we will express the tadpole equations
in the Higgs basis and not in the mass eigenstate basis as the expressions
are simpler. The mass eigenstate basis for $h$ and $H$ and the Higgs
basis is related by an orthogonal transformation parametrized by \citep{Branco:2011iw}:
\begin{align}
 & O^{\phi}=\left(\begin{array}{cc}
c_{\alpha} & s_{\alpha}\\
-s_{\alpha} & c_{\alpha}
\end{array}\right)\,,\,\,\phi_{i}^{mass}=O_{ij}^{\phi}\phi_{j}^{Higgs}\,,\nonumber \\
 & \phi_{i}^{Higgs}=\left(h,H\right)_{i}\,,\label{eq:higgs mix}
\end{align}
where $s_{\alpha}$ and $c_{\alpha}$ are sine and cosine functions
of a mixing angle $\alpha$, respectively. In general, we would have
$3\times3$ mixing matrix, but the imposed CP symmetry on the potential
does not allow $A$ to mix with $h$ and $H$ at tree level. Then the
tadpole functions in different bases are related by:
\begin{align}
T_{h} & =c_{\alpha}T_{h_{\left(m\right)}}-s_{\alpha}T_{H_{\left(m\right)}}\,,\,\,T_{H}=c_{\alpha}T_{H_{\left(m\right)}}+s_{\alpha}T_{h_{\left(m\right)}}\,,\nonumber \\
T_{A} & =T_{A_{\left(m\right)}}\,,
\end{align}
where we added the $m$ in the subscript to indicate that the fields
are in the mass eigenstates. At tree level, the tadpole functions
are:
\begin{align}
\hat{T}_{h}^{\left[0\right]} & =\frac{\delta\Gamma_{\,\,\text{Higgs}}^{\left[0\right]}}{\delta h}=-v\left(m_{11}^{2}+\frac{1}{2}\lambda_{1}v^{2}\right)\,,\nonumber \\
\hat{T}_{H}^{\left[0\right]} & =\frac{\delta\Gamma_{\,\,\text{Higgs}}^{\left[0\right]}}{\delta H}=v\left(m_{12}^{2}-\frac{1}{2}v^{2}\lambda_{6}\right)\,,\nonumber \\
\hat{T}_{A}^{\left[0\right]} & =\frac{\delta\Gamma_{\,\,\text{Higgs}}^{\left[0\right]}}{\delta A}=0\,.\label{eq:Tadpoles CP conserving}
\end{align}
We see that the third tadpole function is already zero in the CP conserving
case. We require the tadpole conditions to hold for all loop levels: 
\begin{equation}
\hat{T}_{h}^{\left[i\right]}=\hat{T}_{H}^{\left[i\right]}=\hat{T}_{A}^{\left[i\right]}=0\,.\label{eq:Tadpole condition}
\end{equation}
The tree level tadpole conditions $\hat{T}_{h}^{\left[0\right]}=\hat{T}_{H}^{\left[0\right]}=\hat{T}_{A_{0}}^{\left[0\right]}=0$
give:
\begin{equation}
m_{11}^{2}=-\frac{1}{2}\lambda_{1}v^{2}\quad\text{and}\quad m_{12}^{2}=\frac{1}{2}\lambda_{6}v^{2}\,.\label{eq:minimum}
\end{equation}

Now we require the tadpole conditions \prettyref{eq:Tadpole condition}
for tree and one loop level together: 
\begin{equation}
\hat{T}^{\left[0\right]}=0\,,\quad\left(T^{\left[1\right]}+\delta\hat{T}^{\left[1\right]}\right)\Big|_{\hat{T}^{\left[0\right]}=0}=0\,,\label{eq:1loop tadpole condition}
\end{equation}
where we indicate in the second equation that we use the relations
from the first condition at the loop order after algebraically deriving
counterterms from the multiplicative constants shown in \prettyref{eq:rescales}.
The one loop tadpole counterterms evaluated at $\hat{T}^{\left[0\right]}=0$
for the CP conserving case then are:
\begin{align}
\delta\hat{T}_{h}^{\left[1\right]} & =\frac{1}{2}\lambda_{1}v^{3}\left(2\delta_{m11}-\delta_{\lambda1}-2\delta_{v}\right)\,,\nonumber \\
\delta\hat{T}_{H}^{\left[1\right]} & =\frac{1}{2}\lambda_{6}v^{3}\left(2\delta_{m12}-\delta_{\lambda6}-2\delta_{v}\right)\,,\nonumber \\
\delta\hat{T}_{A_{0}}^{\left[1\right]} & =0\,.
\end{align}
As $v$ is defined dynamically by \prettyref{eq:minimum}, it isn't
an independent parameter of the theory. This means that one of the
counterterms $\delta_{m11}$, $\delta_{\lambda1}$, $\delta_{v}$ is redundant.
This is because we didn't yet choose which parameter is used over
which from the tree level minimum condition \prettyref{eq:minimum}.
One of the choices is treating $\lambda_{1}$ and $v$ as the independent
ones so that the shift of $m_{11}$ is given by:
\begin{equation}
\delta_{m11}=\frac{1}{2}\delta_{\lambda1}\,.
\end{equation}
Then the shift of the VEV yields the one loop tadpole counterterms,
evaluated at $\hat{T}^{\left[0\right]}=0$ :
\begin{equation}
\delta\hat{T}_{h}^{\left[1\right]}=-\lambda_{1}v^{3}\delta_{v}\,,\label{eq:beta h}
\end{equation}
\begin{equation}
\delta\hat{T}_{H}^{\left[1\right]}=\frac{1}{2}\lambda_{6}v^{3}\left(2\delta_{m12}-\delta_{\lambda6}-2\delta_{v}\right)\,.
\end{equation}
The one loop tadpole conditions \prettyref{eq:1loop tadpole condition}
give:
\begin{equation}
\delta_{v}=\frac{1}{\lambda_{1}v^{3}}T_{h}^{\left[1\right]}\,,\label{eq:VEV renormalization constant}
\end{equation}
\begin{equation}
\left(\delta_{m12}-\frac{1}{2}\delta_{\lambda6}\right)=\frac{1}{v^{3}}\left(\frac{1}{\lambda_{1}}T_{h}^{\left[1\right]}-\frac{1}{\lambda_{6}}T_{H}^{\left[1\right]}\right)\,.
\end{equation}
The $v$ now stands for a loop renormalized VEV or the ``proper VEV''
as in \citep{Fleischer:1980ub}. So far, the construction is similar
to the $\beta_{t}$ scheme of \citep{Actis:2006ra}, "scheme 3" in \citep{Denner:2016etu}
or \citep{Fleischer:1980ub} of the SM, but without the proper relation
of the VEV to the mass terms, it is not yet complete. To complete
it as in \citep{Fleischer:1980ub,Actis:2006ra,Denner:2016etu}, one
identifies the bare mass parameters arising from the proper VEV, rather
than $v_{0}$, as also noted in \citep{Krause:2016gkg,Krause:2016oke,Krause:2017mal,Actis:2006ra,Denner:2016etu,Fleischer:1980ub}.
The idea is to avoid the inclusion of the gauge dependence coming from
$\delta_{v}$ into the definition of the mass counterterm $\delta_{m}$
as will be shown in the next sections. 

\section{Yukawa sector\label{sec:Yukawa-sector}}

The GN model adds a single sterile neutrino $N_{0}$ to the general
2HDM. This sterile neutrino is a gauge singlet under all gauge groups
of the SM and has a Majorana mass term $M_{0}$. To write the Yukawa
couplings, we start in the flavour basis, in which the Yukawa coupling
of the charged fermions to the first Higgs doublet in the Higgs basis
is diagonal. Then the general Yukawa couplings for neutrinos can be seen as two three-vectors $Y^{1}$
and $Y^{2}$. The neutrino Yukawa Lagrangian together with the Majorana
mass term then is written as: 
\begin{equation}
\mathcal{L}_{\text{Yuk}}=-Y_{i}^{1}\nu_{0i}^{F}N_{0}H_{01}-Y_{i}^{2}\nu_{0i}^{F}N_{0}H_{02}-\frac{1}{2}M_{0}N_{0}N_{0}+h.c.\label{eq:Yukawa starting}
\end{equation}
where the superscript $F$ means the flavour basis in which the index
$i=e,\mu,\tau\equiv1,2,3$. The Yukawa couplings $Y_{i}^{1}$ and
$Y_{i}^{2}$ give in general 6 complex parameters and $M_{0}$ gives
1 complex parameter. We absorb four phases into the $\nu_{0i}^{F}$
and $N_{0}$ to get $Y_{i}^{1},M_{0}\in\mathbb{R}$. By a singular
value decomposition, we can parametrize the Yukawa couplings with
only four real parameters: 
\begin{equation}
d_{0},y_{0}\in\mathbb{R}\,,\,\,d_{0}^{\prime}\in\mathbb{C}\,,\label{eq:Yuk parametrizations-1}
\end{equation}
absorbing the other degrees of freedom into the Unitary mixing matrix.
To make the parametrization easy, we decompose it into subsequent
orthogonal rotations $O$ and phase shifts $U$, so that $O^{23}$
produces zero in the second position of $Y^{1}$ ($O_{2j}^{23}Y_{j}^{1}=0$),
$O^{13}$ in the first ($O_{1k}^{13}O_{kj}^{23}Y_{j}^{1}=0$). $U^{\sigma}$
adjusts the phase of the first element of $Y^{2}$ to match it with the phase
of the second element ($\arg(U_{1l}^{\alpha}O_{lk}^{13}O_{kj}^{23}Y_{j}^{2})=\arg(U_{2l}^{\alpha}O_{lk}^{13}O_{kj}^{23}Y_{j}^{2}$),
$O^{12}$ makes the first element of $Y^{2}$ zero ($O_{1m}^{12}U_{ml}^{\alpha}O_{lk}^{13}O_{kj}^{23}Y_{j}^{2}=0$)
and $U^{\rho}$ adjust the phase so that the second element of $Y^{2}$
is real ($U_{2n}^{\beta}O_{nm}^{12}U_{ml}^{\alpha}O_{lk}^{13}O_{kj}^{23}Y_{j}^{2}\in\mathbb{R}$).
Writing $V=\,U^{\beta}O^{12}U^{\alpha}O^{13}O^{23}$, the basis choice
is summarised as:
\begin{align}
 & V_{1j}Y_{j}^{1}=0\,,\quad V_{2j}Y_{j}^{1}=0\,,\quad V_{3j}Y_{j}^{1}=y_{0}\,,\,\nonumber \\
 & V_{1j}Y_{j}^{2}=0\,,\quad V_{2j}Y_{j}^{2}=d_{0}\,,\quad V_{2j}Y_{j}^{2}=d_{0}^{\prime}\,,\nonumber \\
 & d_{0},y_{0}\in\mathbb{R}\,,\quad d_{0}^{\prime}\in\mathbb{C}\,.\label{eq:parametrization}
\end{align}
Note that we are still free to adjust the phase of the first row of
$V$. To combine these rotations with the seesaw transformation, we
combine all neutrinos to a single vector:
\begin{equation}
\nu_{0i}^{F}=\left(\nu_{0e},\nu_{0\mu},\nu_{0\tau},N_{0}\right)_{i}\,.
\end{equation}
As we work in the Higgs basis, only the first Higgs doublet gets the
VEV. With the parametrization \prettyref{eq:parametrization}, the
seesaw transformation acts on the third and fourth component yielding
the whole $4\times4$ mixing matrix: 
\begin{equation}
U=U^{34}V=U^{34}U^{\beta}O^{12}U^{\alpha}O^{13}O^{23}\,\label{eq:mixing PMNS}
\end{equation}
and the relation between the mass eigenstate and the flavour basis
becomes:
\begin{equation}
\nu_{0i}^{mass}=U_{ij}^{*}\nu_{0j}^{F}\,.\label{eq:neutrino mass to flavour}
\end{equation}
All the parametrization of neutrino mixing matrix is summarized in
\prettyref{app:Parametrizations,-assumptions-an}. 

In order to see the differences in the mass terms between the 
tadpole schemes, we first do the usual construction like in, e.g. \citep{Denner:2006ic}, and then
modify it according to the discussion at the end of \prettyref{sec:Scalar-sector-and}.
After the electroweak symmetry breaking, the seesaw mechanism
yields two bare mass eigenvalues $m_{03}$ and $m_{04}$ that
have the relations: 
\begin{equation}
M_{0}=m_{04}-m_{03}\quad\text{and}\quad y_{0}^{2}v_{0}^{2}=2m_{03}m_{04}\,.\label{eq:seesaw params bare}
\end{equation}
The seesaw parameters are expressed in terms of masses:
\begin{equation}
s_{034}^{2}=\frac{m_{03}}{m_{04}+m_{03}}\quad\text{and}\quad c_{034}^{2}=\frac{m_{04}}{m_{04}+m_{03}}\,.\label{eq:seesaw angles bare}
\end{equation}
Note that as long as we stay at tree level, $v_0=v$.
In this basis we have four neutrino states $\nu_{0i}$, where $\nu_{01}$
and $\nu_{02}$ have zero mass, but $\nu_{02}$ is distinguished from
$\nu_{01}$ by its interaction with the second Higgs doublet, i.e.
$\nu_{01}$ does not couple to any of the Higgses. By applying the
rotation \prettyref{eq:neutrino mass to flavour} in \prettyref{eq:Yukawa starting},
using the parametrizations of \prettyref{eq:parametrization}, \prettyref{eq:seesaw params bare},
\prettyref{eq:seesaw angles bare} and insering the explicit Higgs
basis \prettyref{eq:Higgs basis}, we write the Yukawa Lagrangian
part that includes only neutral scalar fields together with the Majorana
mass terms:
\begin{widetext}
\begin{align}
\mathcal{L}_{\text{Yuk}} & =-\frac{1}{2}m_{03}\,\nu_{03}\nu_{03}-\frac{1}{2}m_{04}\,\nu_{04}\nu_{04}-\frac{1}{\sqrt{2}}d_{0}\left(H_{0}+iA_{0}\right)\nu_{02}\left(-is_{034}\nu_{03}+c_{034}\nu_{04}\right)\nonumber \\
 & -\frac{1}{\sqrt{2}}\left[y_{0}\left(h_{0}+i\chi_{Z0}\right)+d_{0}^{\prime}\,\left(H_{0}+iA_{0}\right)\right]\times\left[c_{034}s_{034}\nu_{03}\nu_{03}+i\left(c_{034}^{2}-s_{34}^{2}\right)\nu_{03}\nu_{04}+c_{034}s_{034}\nu_{04}\nu_{04}\right]\nonumber \\
 & +h.c.\,.\label{eq:Neutrino Yukawa-1}
\end{align}
%
We straightforwardly apply the multiplicative renormalization constants,
\prettyref{eq:rescales}, for all the parameters and fields. The tree
level renormalized effective action is then written in the same way
as the bare Lagrangian, except that the parameters and fields are
the renormalized ones:
\begin{align}
\hat{\Gamma}_{\,\,\,\text{Yuk}}^{\left[0\right]} & =-\frac{1}{2}m_{3}\,\nu_{3}\nu_{3}-\frac{1}{2}m_{4}\,\nu_{4}\nu_{4}-\frac{1}{\sqrt{2}}d\left(H+iA\right)\nu_{2}\left(-is_{34}\nu_{3}+c_{34}\nu_{4}\right)\nonumber \\
 & -\frac{1}{\sqrt{2}}\left[y\left(h+i\chi_{Z}\right)+d^{\prime}\,\left(H+iA\right)\right]\times\left[c_{34}s_{34}\nu_{3}\nu_{3}+i\left(c_{34}^{2}-s_{34}^{2}\right)\nu_{3}\nu_{4}+c_{34}s_{34}\nu_{4}\nu_{4}\right]+\hcs\,,
\label{eq:Neutrino Yukawa}
\end{align}
\end{widetext}
where:
\begin{equation}
M=m_{4}-m_{3}\,,\quad y^{2}v^{2}=2m_{3}m_{4}\,.\label{eq:seesaw params}
\end{equation}
\begin{equation}
s_{34}^{2}=\frac{m_{3}}{m_{4}+m_{3}}\,,\quad c_{34}^{2}=\frac{m_{4}}{m_{4}+m_{3}}\,.\label{eq:seesaw angles}
\end{equation}
Having \prettyref{eq:seesaw params bare} and \prettyref{eq:seesaw angles bare}
for the bare theory and \prettyref{eq:seesaw params} and \prettyref{eq:seesaw angles}
for the renormalized one gives us the relations between the renormalization
constants:
\begin{equation}
\delta_{m3}+\delta_{m4}=2\left(\delta_{v}+\delta_{y}\right)\,,\label{eq:mass counterterms 1}
\end{equation}
\begin{equation}
m_{4}\delta_{m4}-m_{3}\delta_{m3}=\left(m_{4}-m_{3}\right)\delta_{M}\,.\label{eq:mass countertems2}
\end{equation}
The mass renormalization constants are fixed by the CMS condition
\citep{Dudenas:2017}:
\begin{equation}
\delta_{mi}=\frac{1}{2}\left(\Sigma_{\nu_{i}\nu_{i}}+\Sigma_{\bar{\nu}_{i}\bar{\nu}_{i}}
+\Sigma_{\nu_{i}\bar{\nu}_{i}}+\Sigma_{\bar{\nu}_{i}\nu_{i}}\right)\Big|_{p^{2}=m_{i}^{2}}
\,,\text{ }m_i\neq0\,,\label{eq:mass counterterm}
\end{equation}
which is nothing more than the usual expression for the OS renormalized
mass counterterm (as in \citep{Denner:1991kt}) extended to the complex
domain and written in Weyl spinor formalism. The CMS condition gives
us the renormalized mass parameters gauge independent, however from
\prettyref{eq:mass counterterms 1} we see that the mass counterterm 
has the $\delta_{v}$ contribution,
which is gauge dependent. Hence in this way the bare masses become
gauge dependent as well. 

Recalling the discussion at the end of \prettyref{sec:Scalar-sector-and}:
to define the gauge invariant mass counterterm we need to identify
the bare mass with the proper VEV \citep{Fleischer:1980ub}. Thus
the bare relation \prettyref{eq:seesaw params bare} is modified to:
\begin{equation}
M_{0}=m_{04}^{\prime}-m_{03}^{\prime}\,,\quad y_{0}^{2}v^{2}=2m_{04}^{\prime}m_{03}^{\prime}\,.\label{eq:seesaw params bare-1}
\end{equation}
so that there is no $\delta_{v}$ in the definition of $\delta_{m}$s.
From $v_{0}=v\left(1+\delta_{v}\right)$ and comparing \prettyref{eq:seesaw params bare}
with \prettyref{eq:seesaw params bare-1}, we get the relationship
between primed (FJ scheme) and unprimed (usual tadpole scheme) mass
parameters: 
\begin{equation}
m_{0i}=m_{0i}^{\prime}+\Delta_{0}\,,\quad\Delta_{0}=2\frac{m_{04}^{\prime}m_{03}^{\prime}\delta_{v}}{m_{04}^{\prime}+m_{03}^{\prime}}\,,i=3,4\,.\label{eq:mass difference}
\end{equation}
As the seesaw mixing parameters depend on the masses, they are shifted
as well:
\begin{align}
s_{034}^{2} & \to s_{034}^{2}+2\delta_{v}c_{034}^{2}s_{034}^{2}\left(c_{034}^{2}-s_{034}^{2}\right)\,,\nonumber \\
c_{034}^{2} & \to c_{034}^{2}-2\delta_{v}c_{034}^{2}s_{034}^{2}\left(c_{034}^{2}-s_{034}^{2}\right)\,.
\end{align}
However, these shifts of the mixing parameters become relevant only
at higher loops than one, so we can drop them from our one loop expressions.
At one loop level, everything is the same as in \prettyref{eq:Neutrino Yukawa-1},
except that the bare mass term Lagrangian for neutrinos becomes:
\begin{equation}
\mathcal{L}_{mass}=-\frac{1}{2}\left(m_{03}^{\prime}+\Delta_{0}\right)\,\nu_{03}\nu_{03}-\frac{1}{2}\left(m_{04}^{\prime}+\Delta_{0}\right)\,\nu_{04}\nu_{04}\,.
\end{equation}
Starting from this bare Lagrangian, \prettyref{eq:mass counterterm}
is modified to: 
\begin{align}
\delta_{mi}^{\prime}=
\frac{1}{2}\left(\Sigma_{\nu_{i}\nu_{i}}+\Sigma_{\bar{\nu}_{i}\bar{\nu}_{i}}
                  +\Sigma_{\nu_{i}\bar{\nu}_{i}}+\Sigma_{\bar{\nu}_{i}\nu_{i}}
             \right)
& \Big|_{p^{2}=m_{i}^{2}}-\frac{\Delta}{m_{i}}\,,
\nonumber\\
\text{ for }m_i\neq \, & \, 0\,,\label{eq:CMS modified}
\end{align}
where:
\begin{equation}
\Delta=2\frac{m_{3}m_{4}\delta_{v}}{m_{4}+m_{3}}\,\label{eq:pinch part}
\end{equation}
is defined with the renormalized masses $m_{3}$ and $m_{4}$. We
see that $\Delta$ is the same for $\nu_{3}$ and $\nu_{4}$.
To check if $\Delta$ cancels the gauge invariance, we analytically
calculate the gauge dependent parts of \prettyref{eq:mass counterterm}
and \prettyref{eq:pinch part} for $\nu_{3}$ and $\nu_{4}$. Note
that in both tadpole schemes the renormalized masses are the same
CMS masses, while the bare masses $m_{0i}$ differ from $m_{0i}^{\prime}$
by $\Delta_{0}$ as in \prettyref{eq:mass difference}.

\section{Arriving at the expressions for renormalization constants\label{sec:Ariving-at-the}}

We use FeynArts \citep{Hahn:2000kx} and FormCalc \citep{Hahn:1998yk}
to arrive at one loop expressions for self energies and tadpoles.
For making the FeynArts model file we found the SARAH \citep{Staub:2008uz}
package to be useful, which allows to quickly generate a model file
from an input of the Lagrangian in terms of Weyl spinors and scalars
in the user specified gauge group representations. It also has some
built in functions to check the consistency of the model. We choose
the Higgs basis by simply putting the VEV of the second Higgs doublet
to zero in the input file. We leave all the other parameters arbitrary
for generating the FeynArts model file and make replacement rules
for the FeynArts model file parameters to implement our parametrization
afterwards. As we work at the one loop level, tree level relations
to simplify one loop diagrams can be used. As discussed in \prettyref{sec:Definitions-and-the},
the CMS keeps the algebraic structure of the bare theory. This means
that for the algebraic simplifications, all the properties and the
relations of bare parameters can be used for the renormalized parameters
in the CMS as well. Hence we can implement these properties and relations
into the assumptions of the ``Mathematica'' file in which we do
these simplifications. Then the results can be consistently continued
to the complex domain afterwards. In the following subsection we show how
we implemented the parametrizations into the FeynArts model file and
the assumptions for the bare parameters that carry over to the algebraic
one loop simplifications. Then we present the results that we got for the
gauge dependent terms in mass and tadpole renormalization. 

\subsection{Getting FeynArts model file \label{subsec:Getting-FeynArts-file} } 
\begin{enumerate}
\item We generate a FeynArts model file using Sarah:
\begin{itemize}
\item We take a Sarah model file for a 2HDM, and define 1 additional gauge
singlet like this:\\
\fbox{\begin{minipage}[t]{0.8\linewidth}%
\texttt{\footnotesize FermionFields[[6]] = \{n, 1, conj[nR],0, 1, 1\}}%
\end{minipage}}\\
where the last three entries are the charges under the gauge groups
(singlets under all of them), the second is the number of families,
the first and the third is the name of the field and its component,    
respectively (see \citep{Staub:2008uz}). 
\item We modify the Yukawa Lagrangian of that model file to include the general
Yukawa couplings of neutrinos with the first and the second Higgs
doublet as in \prettyref{eq:Yukawa starting} in a direct analog to
the quark sector and add the Majorana mass term for the sterile neutrino:
\\
\fbox{\begin{minipage}[t]{0.8\linewidth}%
\texttt{\footnotesize LagYukawan = - ( - Yn1 H1.n.l - Yn2 H2.n.l
+ 1/2 M n.n )}%
\end{minipage}}
\item In the definitions for the ``EWSB'' phase, we set the VEV of the second Higgs
doublet to zero to implement the Higgs basis as in \prettyref{eq:Higgs basis}:\\
\fbox{\begin{minipage}[t]{0.8\linewidth}%
\texttt{\footnotesize DEFINITION[EWSB][VEVs]= }~\\
\texttt{\footnotesize \{ \{H10, \{v, 1/Sqrt{[}2{]}\}, \{sigma1, \textbackslash[ImaginaryI]/Sqrt[2]\},\{phi1,
1/Sqrt[2]\}\},}~\\
\texttt{\footnotesize \{H20, \{0, 1/Sqrt{[}2{]}\}, \{sigma2, \textbackslash[ImaginaryI]/Sqrt[2]\},\{phi2,
1/Sqrt[2]\}\} \};}%
\end{minipage}}
\item We leave the definition of mixing between Higgses $h$ and $H$ as in
the 2HDM model, but omit mixings between the pseudoscalars and the
charged scalars as they do not appear in the Higgs basis with CP conserved
potential. 
\item We define an additional mixing matrix for neutrinos in the
 \texttt{\footnotesize DEFINITION[EWSB][MatterSector]},
combining the flavour basis SM neutrinos \texttt{vL} with the sterile neutrino
\texttt{conj(nR)} as:\\
\fbox{\begin{minipage}[t]{0.8\linewidth}%
\texttt{\footnotesize \{\{vL,conj[nR]\}, \{VL,Un\}\} \}}%
\end{minipage}}\\
where the \texttt{VL} is the combined four-vector of the neutrino mass eigenstates
and \texttt{Un} is the mixing matrix $U^{*}$ from \prettyref{eq:neutrino mass to flavour}.
\item We generate the FeynArts model file by the Sarah command \texttt{MakeFeynArts[]}.
\end{itemize}
\item We make modifications to the FeynArts model file:
\begin{itemize}
\item To achieve the parametrization of \prettyref{eq:parametrization}
we make the replacements in the model file for the neutrino-neutrino
- Higgs vertices: 
\begin{equation}
\sum_{j=1}^{3}U_{ij}Y_{j}^{1}\to\left(0,\,0,\,-i\,c_{34}y,\,s_{34}y\right)_{j}\,,\label{eq:first parametrization}
\end{equation}
\begin{equation}
\sum_{j=1}^{3}U_{ij}Y_{j}^{2}\to\left(0,\,d,\,-i\,c_{34}d^{\prime},\,s_{34}d^{\prime}\right)_{j}\,.\label{eq:second parametrization}
\end{equation}
We do not replace the neutrino - electron - scalar vertices, hence
they depend on $Y^{1}$ and $Y^{2}$ instead of the $y,d$ and $d'$\textbf{
}parameters in the model file. We leave them general, because it is
easier to make algebraic simplifications of amplitudes in the general
couplings for these vertices. After the expressions are simple enough,
we invert \prettyref{eq:first parametrization} and \prettyref{eq:second parametrization}
to express $Y^{1}$ and $Y^{2}$ in terms of $U\,,\,y,d$ and $d'$
in the Mathematica notebook file. 
\end{itemize}
\end{enumerate}
After setting up the FeynArts model file, we generate 1 loop diagrams for
the wanted correlation functions. The parametrizations and relations
of \prettyref{sec:Scalar-sector-and} and \prettyref{sec:Yukawa-sector}
are imposed as replacement rules during the algebraic simplifications
of the expressions. The summary of the parameters and their relations
is given in the \prettyref{app:Parametrizations,-assumptions-an}. 

\subsection{Mass renormalization}

We construct the mass renormalization constants as in \prettyref{eq:mass counterterm}
to isolate the gauge dependent part so that we can later check if
the definition in \prettyref{eq:CMS modified} really cancels it.
The FormCalc output is easy to use in Weyl spinor notation as the
spinor products in the result of the amplitude appear in ``WeylChains''.
By collecting terms near those ``WeylChains'' we can take separately
all four components presented in \prettyref{eq:selfs}. The structure
of the correction to a propagator is:
\begin{equation}
\left\langle \nu_{i}^{\,}\nu_{i}^{\,}\right\rangle \Gamma_{\nu_{i}\nu_{i}}+\left\langle \bar{\nu}_{i}\bar{\nu}_{i}\right\rangle \Gamma_{\bar{\nu}_{i}\bar{\nu}_{i}}+\left\langle \nu_{i}p\sigma\bar{\nu}_{i}\right\rangle \Sigma_{\nu_{i}\bar{\nu}_{i}}+\left\langle \bar{\nu}_{i}p\bar{\sigma}\nu_{i}\right\rangle \Sigma_{\bar{\nu}_{i}\nu_{i}}.
\end{equation}
For Majorana particles only two of the scalar self energies are independent,
since $\Sigma_{\nu\bar{\nu}}$ is the same as $\Sigma_{\bar{\nu}\nu}$
and $\Gamma_{\nu\nu}$ is related to $\Gamma_{\bar{\nu}\bar{\nu}}$.
At one loop, this relation is just the hermitian conjugation of couplings
that enter the loop functions. 

To make algebra simplifications easier and faster we separate different
one loop contributions to self energies according to the particles
that appear in the loop. Those contributions are from the neutral
Higgs scalars, the charged scalar Higgs, the neutral Goldstone boson,
the charged Goldstone boson, the W boson and the Z boson. We label
them as $\Sigma^{H0}$, $\Sigma^{H+}$, $\Sigma^{\chi0}$, $\Sigma^{\chi+}$, $\Sigma^{W}$ and $\Sigma^{Z}$,
respectively. Note that the $\Sigma$s are the dimensionless one loop self energy functions
defined in \prettyref{eq:selfs-1}. Analogously, we write the dimensionful self energies
as $\Gamma^{H0}_{\phi_1\phi_2}$, $\Gamma^{H+}_{\phi_1\phi_2}$, etc\dots 
Naturally, $\Gamma^{H0}_{\nu_i\nu_j}$ and $\Gamma^{H+}_{\nu_i\nu_j}$ do not
depend on any gauge parameter. As the first results of the calculations
give us: 
\begin{equation}
\Gamma_{\nu_{1}\nu_{1}}^{\left[1\right]}=0\text{ and }\Gamma_{\nu_{2}\nu_{2}}^{\left[1\right]}=\Gamma_{\nu_{2}\nu_{2}}^{H0}\,.\label{eq:nu1 and nu2}
\end{equation}
Note that $\nu_{2}$ and $\nu_{1}$ do not have mass renormalization
constants coming from \prettyref{eq:rescales}, since they do not have
bare mass parameters. The non vanishing contribution for the mass
of $\nu_{2}$ is gauge independent and finite. This is a good first
crosscheck to see that the implementation of the model gives us expected
results. 

We are interested in the gauge dependent part of $\delta_{m3}$
and $\delta_{m4}$, so we are interested only in $\Sigma^{\chi0}$, $\Sigma^{\chi+}$, $\Sigma^{W}$ and $\Sigma^{Z}$.
$\xi_{W}$ will appear only in $\Sigma^{\chi+}$ and $\Sigma^{W}$
and $\xi_{Z}$ only in $\Sigma^{\chi0}$ and $\Sigma^{Z}$.
As one can check, the charged loop for mass-like terms vanishes:
\begin{equation}
\Gamma_{\nu_{3}\nu_{3}}^{W}=\Gamma_{\nu_{3}^{\dagger}\nu_{3}^{\dagger}}^{W}=0\,.
\end{equation}
Hence the potentially $\xi_{W}$ dependent contribution for $m_{3}\delta_{m3}$
is:
\begin{equation}
\frac{1}{2}\left(\Gamma_{\nu_{3}\nu_{3}}^{\chi+}+\Gamma_{\nu_{3}^{\dagger}\nu_{3}^{\dagger}}^{\chi+}\right)+m_{3}\Sigma_{\nu_{3}\nu_{3}^{\dagger}}^{W}+m_{3}\Sigma_{\nu_{3}\nu_{3}^{\dagger}}^{\chi+}\,.\label{eq:gauge dependent starting point}
\end{equation}
After some effort (see the \prettyref{app:appendix1}), we arrive
at the $\xi_{W}$ dependent part of the mass counterterm (recall \prettyref{eq:Gauge dependence notation}):
\begin{equation}
m_{3}\delta_{m_{3}\xi_{W}}=\frac{m_{3}m_{4}}{\left(m_{3}+m_{4}\right)}\frac{g_{e}^{2}}{16\pi^{2}m_{Z}^{2}s_{2W}^{2}}2\,A_{0}\left(m_{W}^{2}\xi_{W}\right)\,,\label{eq:w dependent}
\end{equation}
where $s_{2W}\equiv2s_{W}c_{W}$ is the sine of a double Weinberg
angle \prettyref{eq:SM relations}. For calculating $\delta_{m_{3}\xi_{Z}}$
one should note that $\Gamma_{\nu_{3}\nu_{3}}^{Z}\neq0$. Apart from
that, everything is analogous to the $\xi_{W}$ case. At the end the
full gauge dependence of the neutrino mass counterterms is:
\begin{widetext}
\begin{equation}
m_{3}\delta_{m_{3}\xi}=m_{4}\delta_{m_{4}\xi}=\frac{m_{3}m_{4}}{\left(m_{3}+m_{4}\right)}\frac{g_{e}^{2}}{16\pi^{2}m_{Z}^{2}s_{2W}^{2}}\left[A_{0}\left(m_{Z}^{2}\xi_{Z}\right)+2\,A_{0}\left(m_{W}^{2}\xi_{W}\right)\right]\,.\label{eq:deltam gauge}
\end{equation}
\end{widetext}

\subsection{VEV renormalization}

When separating the gauge parameter dependent part of $T_{h}^{\left[1\right]}$
we first observe that tadpoles with physical Higgs bosons and fermions
in the loop do not have any gauge dependence. The gauge dependent
part of loops with gauge bosons and ghosts exactly cancel when these
contributions are summed up. Hence the only gauge dependent terms
in the tadpole contributions are the tadpoles with Goldstone bosons
in the loops, which are: 
\begin{equation}
T_{h\xi}^{\left[1\right]}=\frac{\lambda_{1}v}{32\pi^{2}}\left[A_{0}\left(m_{Z}^{2}\xi_{Z}\right)+2\,A_{0}\left(m_{W}^{2}\xi_{W}\right)\right]\,.\label{eq:Th gauge}
\end{equation}
This is exactly the same term that we would get for the Higgs tadpole
in the SM. This again shows the convenience of the Higgs basis in
the tadpole equations. From \prettyref{eq:VEV renormalization constant}
and \prettyref{eq:pinch part} we have:
\begin{equation}
\Delta_{\xi}=\frac{m_{3}m_{4}}{\left(m_{3}+m_{4}\right)}\frac{1}{16\pi^{2}v^{2}}\left[A_{0}\left(m_{Z}^{2}\xi_{Z}\right)+2\,A_{0}\left(m_{W}^{2}\xi_{W}\right)\right]\,,\label{eq:gauge delta}
\end{equation}
which, inserting the SM relations of \prettyref{eq:SM relations}
gives exactly the same result as \prettyref{eq:deltam gauge}. 

\clearpage

\section{Discussion and conclusions\label{sec:Gauge-dependence-of-1}}

We analytically checked in the CMS or the OS scheme that the gauge
dependent term of the mass counterterms for the neutrinos of the GN
model comes only from the tadpole contributions, \prettyref{eq:deltam gauge},
as suggested in \citep{Liebler:2010bi}. Using multiplicative renormalization
constants and the relations between them, shown in \prettyref{eq:mass counterterms 1}
and \prettyref{eq:mass countertems2}, we present how the gauge dependence
of neutrino mass counterterms can be seen as a contribution coming
from $\delta_{v}$, the renormalization constant of the VEV in the
usual tadpole renormalization (for example \citep{Denner:1991kt}).
We also get that this tadpole contribution is the same for both neutrino
counterterms:
\begin{equation}
m_{3}\delta_{m_{3}\xi}=m_{4}\delta_{m_{4}\xi}=\Delta_{\xi}\,.\label{eq:gauge dependent terms}
\end{equation}
This is one of the features of the GN model: the single sterile 
neutrino leads to the single value of
the Yukawa coupling $y$ to the first Higgs doublet in the Higgs basis.
This single value is coupled to the VEV, hence only the single value
$\Delta$, related to the VEV shift $\delta_{v}$, is possible for
the neutrino mass counterterms in this setup. 

The alternative tadpole scheme, or the FJ scheme \citep{Fleischer:1980ub},
consistently omits this gauge dependence from the mass renormalization
constants by identifying the bare masses with the proper VEV. Following
this scheme, we modify the definition of the mass counterterms to
include this tadpole contribution in \prettyref{eq:CMS modified}.
This definition now exactly cancels the gauge dependent contribution
as can be seen from \prettyref{eq:gauge dependent terms}. The factor
$\Delta$ gives the same contribution for the mass counterterms as
if we would add the contribution of diagrams with tadpoles connected
to the propagators as in \citep{Liebler:2010bi}. The fact that the
procedures of \citep{Fleischer:1980ub} works for the seesaw neutrinos
just in the same way as with the Dirac particles is explained by the
fact that only the Dirac mass ($\sim m_{3}m_{4}$ from \prettyref{eq:seesaw params bare})
is directly related to the VEV. The other crosscheck is that the result
of \prettyref{eq:gauge dependent terms}, using \prettyref{eq:mass countertems2},
gives 
\begin{equation}
\delta_{M\xi}=0\,,
\end{equation}
or in other words, the Majorana mass term $M$, does not acquire gauge
dependence in any of these schemes. This again confirms the statement
that the Majorana mass term of the sterile part of the neutrino doesn't
affect the application of the FJ scheme for mass counterterms for
the neutrinos. Hence using the FJ scheme is straightforwardly
applicable in the GN model. 
\begin{acknowledgments}
The authors thank the Lithuanian Academy of Sciences for the support
(the project DaFi2017).
\end{acknowledgments}

\appendix

\section{Parametrizations, assumptions and relations\label{app:Parametrizations,-assumptions-an}}

Here we collect all parameters and relations used in our 1 loop
calculations. The assumption that some bare parameter $p_{0}$ is
real, is reflected in the renormalized theory in the sense of \prettyref{eq:hc star}.
In the FormCalc output for one loop corrections for masses, we implement
this assumption by the replacement rule $p^{\dagger}\to p$, for $p_{0}\in\mathbb{R}$. 

\subsection{Scalar sector and the SM relations}

The assumptions of CP conservation of the Higgs potential give:
\begin{equation}
m_{0ij}^{2},\lambda_{0k}\in\mathbb{R}\,;\,\,i,j=1,2\,,\,k=1,...,7.\label{eq:CP conserving-1}
\end{equation}
The minimum conditions are: 
\begin{equation}
m_{11}^{2}=-\frac{1}{2}\lambda_{1}v^{2}\quad\text{and}\quad m_{12}^{2}=\frac{1}{2}\lambda_{6}v^{2}\,.\label{eq:minimum-1}
\end{equation}
The Higgs basis is given by: 
\begin{equation}
H_{1}=\left(\begin{array}{c}
\chi_{0W}^{+}\\
\frac{1}{\sqrt{2}}\left(v+h+i\chi_{Z}\right)
\end{array}\right)\,,\,\,H_{2}=\left(\begin{array}{c}
H_{0}^{+}\\
\frac{1}{\sqrt{2}}\left(H+iA\right)
\end{array}\right)\,.\label{eq:Higgs basis-2}
\end{equation}
The mixing matrix for scalars is only between $h$ and $H$:
\begin{align}
 & O^{\phi}=\left(\begin{array}{cc}
c_{\alpha} & s_{\alpha}\\
-s_{\alpha} & c_{\alpha}
\end{array}\right)\,,\,\,\phi_{i}^{mass}=O_{ij}^{\phi}\phi_{j}^{Higgs}\,,\nonumber \\
 & \phi_{i}^{Higgs}=\left(h,H\right)_{i}\,,
\end{align}
where $s_{\alpha},c_{\alpha}$ are sine and cosine functions of the
mixing angle $\alpha$. 

The relations of the Electroweak sector are:
\begin{equation}
s_{2W}\equiv2s_{W}c_{W}\,,\,\,m_{Z}=\frac{g_{e}v}{s_{2W}}\,,\,m_{W}=m_{Z}c_{W}\,,\label{eq:SM relations}
\end{equation}
where $s_{W}$ and $c_{W}$are sine and cosine functions of Weinberg
angle. 

\subsection{Yukawa sector}

As the first thing after generating the FeynArts model file we
make the replacements \prettyref{eq:first parametrization} and
\prettyref{eq:second parametrization}:
\begin{align}
\sum_{j=1}^{3}U_{ij}Y_{j}^{1} & \to\left(0,\,0,\,-i\,c_{34}y,\,s_{34}y\right)_{j}\,,\nonumber \\
\sum_{j=1}^{3}U_{ij}Y_{j}^{2} & \to\left(0,\,d,\,-i\,c_{34}d^{\prime},\,s_{34}d^{\prime}\right)_{j}\,.
\end{align}
The parametrization of Yukawa couplings are summarized as:
\begin{align}
 & V_{1j}Y_{j}^{1}=0\,,\quad V_{2j}Y_{j}^{1}=0\,,\quad V_{3j}Y_{j}^{1}=y_{0}\,,\,\nonumber \\
 & V_{1j}Y_{j}^{2}=0\,,\quad V_{2j}Y_{j}^{2}=d_{0}\,,\quad V_{2j}Y_{j}^{2}=d_{0}^{\prime}\,,\nonumber \\
 & d_{0},y_{0}\in\mathbb{R}\,,\quad d_{0}^{\prime}\in\mathbb{C}\,,\label{eq:parametrization-1}
\end{align}
where the neutrino mixing matrix is:
\begin{equation}
U=U^{34}V=U^{34}U^{\beta}O^{12}U^{\alpha}O^{13}O^{23}\,\label{eq:mixing PMNS-1}
\end{equation}
with the relations 
\begin{equation}
\nu_{0i}^{F}=\left(\nu_{0e},\nu_{0\mu},\nu_{0\tau},N_{0}\right)_{i}\,,\,\,\nu_{0i}^{mass}=U_{ij}^{*}\nu_{0j}^{F}\,.\label{eq:neutrino mass to flavour-1}
\end{equation}
The parametrization of the mixing matrix can be written as:
\begin{align}
 & s_{0ij}^{2}+c_{0ij}^{2}=1\,,\,\,s_{0ij},c_{0ij},\sigma_{0},\rho_{0}\in\mathbb{R}\,;\nonumber \\
 & O_{ij}^{AB}=1_{ij}\mbox{ for }\,i,j\neq A,B\,;\nonumber \\
 & O_{AB}^{AB}=-O_{BA}^{AB}=s_{0AB}\,;\,\,O_{AA}^{AB}=O_{BB}^{AB}=c_{0AB}\,;\nonumber \\
 & U_{ij}^{\sigma}=e^{i\sigma_{0}}\mbox{ for }i=j=1;\,U_{ij}^{\sigma}=1_{ij}\mbox{ for }i,j\neq1\,;\nonumber \\
 & U_{ij}^{\rho}=e^{i\rho_{0}}\mbox{ for }i=j=2;\,U_{ij}^{\rho}=1_{ij}\mbox{ for }i,j\neq2\,;\nonumber \\
 & U_{34}^{34}=i\cdot U_{43}^{34}=i\cdot s_{034};\,\,U_{33}^{34}=-i\cdot U_{44}^{34}=-i\cdot c_{034};\nonumber \\
 & U_{ij}^{34}=1_{ij}\mbox{ for }i,j\neq3,4\,.
\end{align}
The seesaw mechanism is realized with: 
\begin{equation}
M_{0}=m_{04}-m_{03}\,,\quad\quad y_{0}^{2}v_{0}^{2}=2m_{03}m_{04}\,,\label{eq:seesaw params bare-2}
\end{equation}
\begin{equation}
s_{034}^{2}=\frac{m_{03}}{m_{04}+m_{03}}\quad\text{and}\quad c_{034}^{2}=\frac{m_{04}}{m_{04}+m_{03}}\,.\label{eq:seesaw angles bare-2}
\end{equation}

\section{\label{app:appendix1}Arriving at \prettyref{eq:w dependent}}

Here we show some intermediate steps for arriving at the gauge
parameter $\xi_{W}$ dependent term for the $\delta_{m3}$ counterterm
shown in \prettyref{eq:w dependent}. We start from \prettyref{eq:gauge dependent starting point}:
\begin{equation}
\frac{1}{2}\left(\Gamma_{\nu_{3}\nu_{3}}^{\chi+}+\Gamma_{\nu_{3}^{\dagger}\nu_{3}^{\dagger}}^{\chi+}\right)+m_{3}\Sigma_{\nu_{3}\nu_{3}^{\dagger}}^{W}+m_{3}\Sigma_{\nu_{3}\nu_{3}^{\dagger}}^{\chi+}\,.\label{eq:initial gauge dependent}
\end{equation}
Let us first look at the loop with the Goldstone boson $\Gamma_{\nu_{3}\nu_{3}}^{\chi+}\left(m_{3}^{2}\right)$.
We set up the model file in FeynArts following the steps in \prettyref{subsec:Getting-FeynArts-file}.
After generating diagrams with FeynArts, creating an amplitude with
FormCalc, implementing the parametrization that is summarised in \prettyref{app:Parametrizations,-assumptions-an}
by the replacement rules, the standard Mathematica ``Simplify''
command should give: 
\begin{widetext}
\begin{align}
\Gamma_{\nu_{3}\nu_{3}}^{\chi+}\left(m_{3}^{2}\right) & =\frac{-\sqrt{m_{3}m_{4}}}{4\sqrt{2}\pi^{2}(m_{3}+m_{4})v}\Big[-m_{\tau}^{2}Y_{\tau}^{1*}c_{13}c_{23}B_{0}\left(m_{3}^{2},m_{W}^{2}\xi_{W},m_{\tau}^{3}\right)\nonumber \\
 & +m_{\mu}^{2}Y_{\mu}^{1*}c_{13}s_{23}B_{0}\left(m_{3}^{2},m_{W}^{2}\xi_{W},m_{\mu}^{3}\right)+m_{e}^{2}Y_{e}^{1*}s_{13}B_{0}\left(m_{3}^{2},m_{W}^{2}\xi_{W},m_{e}^{3}\right)\Big]\,.
\end{align}
Expressing $Y^{1}$ from \prettyref{eq:first parametrization} and
\prettyref{eq:second parametrization} gives:
\begin{align}
\Gamma_{\nu_{3}\nu_{3}}^{\chi+}\left(m_{3}^{2}\right) & =\frac{-y\sqrt{m_{3}m_{4}}}{4\sqrt{2}\pi^{2}(m_{3}+m_{4})v}\Big[m_{\tau}^{2}c_{13}^{2}c_{23}^{2}B_{0}\left(m_{3}^{2},m_{W}^{2}\xi_{W},m_{\tau}^{3}\right)\nonumber \\
 & +m_{\mu}^{2}c_{13}^{2}s_{23}^{2}B_{0}\left(m_{3}^{2},m_{W}^{2}\xi_{W},m_{\mu}^{3}\right)+m_{e}^{2}s_{13}^{2}B_{0}\left(m_{3}^{2},m_{W}^{2}\xi_{W},m_{e}^{3}\right)\Big]\,.
\end{align}
Now we can express $v$ in terms of \prettyref{eq:SM relations} and
$y$ in terms of \prettyref{eq:seesaw params} to get: 
\begin{align}
\Gamma_{\nu_{3}\nu_{3}}^{\chi+}\left(m_{3}^{2}\right) & =\frac{-g_{e}^{2}m_{3}m_{4}}{4\pi^{2}(m_{3}+m_{4})m_{Z}^{2}s_{2W}^{2}}\Big[m_{\tau}^{2}c_{13}^{2}c_{23}^{2}B_{0}\left(m_{3}^{2},m_{W}^{2}\xi_{W},m_{\tau}^{3}\right)\nonumber \\
 & +m_{\mu}^{2}c_{13}^{2}s_{23}^{2}B_{0}\left(m_{3}^{2},m_{W}^{2}\xi_{W},m_{\mu}^{3}\right)+m_{e}^{2}s_{13}^{2}B_{0}\left(m_{3}^{2},m_{W}^{2}\xi_{W},m_{e}^{3}\right)\Big]\,.
\end{align}
The result for $\Gamma_{\nu_{3}^{\dagger}\nu_{3}^{\dagger}}^{\chi+}$
is the same, as it should be, since $\nu_{3}$ is a Majorana fermion
and the couplings can be taken real for the one loop correction, hence
we can write:
\begin{align}
\frac{1}{2}\left(\Gamma_{\nu_{3}\nu_{3}}^{\chi+}+\Gamma_{\nu_{3}^{\dagger}\nu_{3}^{\dagger}}^{\chi+}\right) & =\frac{-g_{e}^{2}m_{3}m_{4}}{4\pi^{2}(m_{3}+m_{4})m_{Z}^{2}s_{2W}^{2}}\times\Big[m_{\tau}^{2}c_{13}^{2}c_{23}^{2}B_{0}\left(m_{3}^{2},m_{W}^{2}\xi_{W},m_{\tau}^{3}\right)\nonumber \\
 & +m_{\mu}^{2}c_{13}^{2}s_{23}^{2}B_{0}\left(m_{3}^{2},m_{W}^{2}\xi_{W},m_{\mu}^{3}\right)+m_{e}^{2}s_{13}^{2}B_{0}\left(m_{3}^{2},m_{W}^{2}\xi_{W},m_{e}^{3}\right)\Big]\,.\label{eq:scal goldstone looop}
\end{align}
We follow exactly the same steps for $\Sigma_{\nu_{3}\nu_{3}^{\dagger}}^{\chi+}$
to get:
\begin{align}
 & \frac{g_{e}^{2}m_{4}}{8\pi^{2}(m_{3}+m_{4})m_{Z}^{2}s_{2W}^{2}}\times\nonumber \\
 & \times\Big[m_{3}^{2}c_{13}^{2}c_{23}^{2}B_{0}\left(m_{3}^{2},m_{W}^{2}\xi_{W},m_{\tau}^{2}\right)+m_{3}^{2}c_{13}^{2}s_{23}^{2}B_{0}\left(m_{3}^{2},m_{W}^{2}\xi_{W},m_{\mu}^{2}\right)+m_{3}^{2}s_{13}^{2}B_{0}\left(m_{3}^{2},m_{W}^{2}\xi_{W},m_{e}^{2}\right)\nonumber \\
 & +m_{\tau}^{2}c_{13}^{2}c_{23}^{2}B_{0}\left(m_{3}^{2},m_{W}^{2}\xi_{W},m_{\tau}^{2}\right)+m_{\mu}^{2}c_{13}^{2}s_{23}^{2}B_{0}\left(m_{3}^{2},m_{W}^{2}\xi_{W},m_{\mu}^{2}\right)+m_{e}^{2}s_{13}^{2}B_{0}\left(m_{3}^{2},m_{W}^{2}\xi_{W},m_{e}^{2}\right)\nonumber \\
 & +m_{3}^{2}c_{13}^{2}c_{23}^{2}B_{1}\left(m_{3}^{2},m_{W}^{2}\xi_{W},m_{\tau}^{2}\right)+m_{3}^{2}c_{13}^{2}s_{23}^{2}B_{1}\left(m_{3}^{2},m_{W}^{2}\xi_{W},m_{\mu}^{2}\right)+m_{3}^{2}s_{13}^{2}B_{1}\left(m_{3}^{2},m_{W}^{2}\xi_{W},m_{e}^{2}\right)\nonumber \\
 & +m_{\tau}^{2}c_{13}^{2}c_{23}^{2}B_{1}\left(m_{3}^{2},m_{W}^{2}\xi_{W},m_{\tau}^{2}\right)+m_{\mu}^{2}c_{13}^{2}s_{23}^{2}B_{1}\left(m_{3}^{2},m_{W}^{2}\xi_{W},m_{\mu}^{2}\right)+m_{e}^{2}s_{13}^{2}B_{1}\left(m_{3}^{2},m_{W}^{2}\xi_{W},m_{e}^{2}\right)\Big]\,.\label{eq:vec goldstone loop}
\end{align}
The loop with the $W$ boson $\Sigma_{\nu_{3}\nu_{3}^{\dagger}}^{W}$
will have gauge invariant contributions from the transverse polarization
of the W boson. These can be dropped out from the expression by formally
differentiating and integrating with respect to $\xi_{W}$ in Mathematica.
Then every step for simplifying the expression is the same as before with the
result: 
\begin{align}
 & \frac{g_{e}^{2}m_{4}}{8\pi^{2}(m_{3}+m_{4})m_{Z}^{2}s_{2W}^{2}}\times\nonumber \\
 & \times\Big[-m_{3}^{2}c_{13}^{2}c_{23}^{2}B_{0}\left(m_{3}^{2},m_{W}^{2}\xi_{W},m_{\tau}^{2}\right)-m_{3}^{2}c_{13}^{2}s_{23}^{2}B_{0}\left(m_{3}^{2},m_{W}^{2}\xi_{W},m_{\mu}^{2}\right)-m_{3}^{2}s_{13}^{2}B_{0}\left(m_{3}^{2},m_{W}^{2}\xi_{W},m_{e}^{2}\right)\nonumber \\
 & +m_{\tau}^{2}c_{13}^{2}c_{23}^{2}B_{0}\left(m_{3}^{2},m_{W}^{2}\xi_{W},m_{\tau}^{2}\right)+m_{\mu}^{2}c_{13}^{2}s_{23}^{2}B_{0}\left(m_{3}^{2},m_{W}^{2}\xi_{W},m_{\mu}^{2}\right)+m_{e}^{2}s_{13}^{2}B_{0}\left(m_{3}^{2},m_{W}^{2}\xi_{W},m_{e}^{2}\right)\nonumber \\
 & -m_{3}^{2}c_{13}^{2}c_{23}^{2}B_{1}\left(m_{3}^{2},m_{W}^{2}\xi_{W},m_{\tau}^{2}\right)-m_{3}^{2}c_{13}^{2}s_{23}^{2}B_{1}\left(m_{3}^{2},m_{W}^{2}\xi_{W},m_{\mu}^{2}\right)-m_{3}^{2}s_{13}^{2}B_{1}\left(m_{3}^{2},m_{W}^{2}\xi_{W},m_{e}^{2}\right)\nonumber \\
 & -m_{\tau}^{2}c_{13}^{2}c_{23}^{2}B_{1}\left(m_{3}^{2},m_{W}^{2}\xi_{W},m_{\tau}^{2}\right)-m_{\mu}^{2}c_{13}^{2}s_{23}^{2}B_{1}\left(m_{3}^{2},m_{W}^{2}\xi_{W},m_{\mu}^{2}\right)-m_{e}^{2}s_{13}^{2}B_{1}\left(m_{3}^{2},m_{W}^{2}\xi_{W},m_{e}^{2}\right)\nonumber \\
 & +c_{13}^{2}c_{23}^{2}A_{0}\left(m_{W}^{2}\xi_{W}\right)+c_{13}^{2}s_{23}^{2}A_{0}\left(m_{W}^{2}\xi_{W}\right)+s_{13}^{2}A_{0}\left(m_{W}^{2}\xi_{W}\right)\Big]\,.\label{eq:w boson loop}
\end{align}
Comparing \prettyref{eq:w boson loop} with \prettyref{eq:vec goldstone loop}
we notice that the first, third and fourth lines of both expressions
cancel and the second line of both equations is the same. Trigonometric
functions near the $A_{0}$ integrals in \prettyref{eq:w boson loop}
sum to one. The sum of \prettyref{eq:w boson loop} and \prettyref{eq:vec goldstone loop}
multiplied by $m_{3}$ then gives 
\begin{align}
 & \frac{g_{e}^{2}m_{4}m_{3}}{4\pi^{2}(m_{3}+m_{4})m_{Z}^{2}s_{2W}^{2}}\times\nonumber \\
 & \times\Big[m_{\tau}^{2}c_{13}^{2}c_{23}^{2}B_{0}\left(m_{3}^{2},m_{W}^{2}\xi_{W},m_{\tau}^{2}\right)+m_{\mu}^{2}c_{13}^{2}s_{23}^{2}B_{0}\left(m_{3}^{2},m_{W}^{2}\xi_{W},m_{\mu}^{2}\right)+m_{e}^{2}s_{13}^{2}B_{0}\left(m_{3}^{2},m_{W}^{2}\xi_{W},m_{e}^{2}\right)\Big]\nonumber \\
 & +\frac{g_{e}^{2}m_{4}m_{3}}{8\pi^{2}(m_{3}+m_{4})m_{Z}^{2}s_{2W}^{2}}A_{0}\left(m_{W}^{2}\xi_{W}\right)\,.
\end{align}
The second line cancels with the contribution of the Goldstone loop
from \prettyref{eq:scal goldstone looop} giving exactly \prettyref{eq:w dependent}.

\end{widetext}

\bibliographystyle{utphys}
\bibliography{references}

\providecommand{\href}[2]{#2}\begingroup\raggedright\begin{thebibliography}{10}

\bibitem{Heeger:2004mp}
K.~M. Heeger, \href{http://dx.doi.org/10.1142/9789812702210_0005}{``{Evidence
  for neutrino mass: A Decade of discovery},''} in {\em {Seesaw mechanism.
  Proceedings, International Conference, SEESAW25, Paris, France, June 10-11,
  2004}}, pp.~65--80.
\newblock 2004.
\newblock
\href{http://arxiv.org/abs/hep-ex/0412032}{{\ttfamily arXiv:hep-ex/0412032
  [hep-ex]}}.
\newblock

\bibitem{Molinaro:2013toa}
E.~Molinaro, ``{Type I Seesaw Mechanism, Lepton Flavour Violation and Higgs
  Decays},'' \href{http://dx.doi.org/10.1088/1742-6596/447/1/012052}{{\em
  J.Phys.Conf.Ser.} {\bfseries 447} (2013) 012052},
\href{http://arxiv.org/abs/1303.5856}{{\ttfamily arXiv:1303.5856 [hep-ph]}}.

\bibitem{Orloff:2005nu}
J.~Orloff, S.~Lavignac, and M.~Cribier, eds., {\em {Seesaw mechanism.
  Proceedings, International Conference, SEESAW25, Paris, France, June 10-11,
  2004}}.
\newblock
2005.
\newblock

\bibitem{Grimus:1989pu}
W.~Grimus and H.~Neufeld, ``{Radiative Neutrino Masses in an SU(2) X U(1)
  Model},''
\href{http://dx.doi.org/10.1016/0550-3213(89)90370-2}{{\em Nucl. Phys.}
  {\bfseries B325} (1989) 18--32}.

\bibitem{Haber:1984rc}
H.~E. Haber and G.~L. Kane, ``{The Search for Supersymmetry: Probing Physics
  Beyond the Standard Model},''
\href{http://dx.doi.org/10.1016/0370-1573(85)90051-1}{{\em Phys. Rept.}
  {\bfseries 117} (1985) 75--263}.

\bibitem{Kim:1986ax}
J.~E. Kim, ``{Light Pseudoscalars, Particle Physics and Cosmology},''
\href{http://dx.doi.org/10.1016/0370-1573(87)90017-2}{{\em Phys. Rept.}
  {\bfseries 150} (1987) 1--177}.

\bibitem{Lee:1973iz}
T.~D. Lee, ``A theory of spontaneous $t$ violation,''
  \href{http://dx.doi.org/10.1103/PhysRevD.8.1226}{{\em Phys. Rev. D}
  {\bfseries 8} (Aug, 1973) 1226--1239}.
  \url{https://link.aps.org/doi/10.1103/PhysRevD.8.1226}.

\bibitem{Peccei:1977hh}
R.~D. Peccei and H.~R. Quinn, ``{CP Conservation in the Presence of
  Instantons},''
\href{http://dx.doi.org/10.1103/PhysRevLett.38.1440}{{\em Phys. Rev. Lett.}
  {\bfseries 38} (1977) 1440--1443}.

\bibitem{Branco:2011iw}
G.~C. Branco, P.~M. Ferreira, L.~Lavoura, M.~N. Rebelo, M.~Sher, and J.~P.
  Silva, ``{Theory and phenomenology of two-Higgs-doublet models},''
  \href{http://dx.doi.org/10.1016/j.physrep.2012.02.002}{{\em Phys. Rept.}
  {\bfseries 516} (2012) 1--102},
\href{http://arxiv.org/abs/1106.0034}{{\ttfamily arXiv:1106.0034 [hep-ph]}}.

\bibitem{Gambino:1999ai}
P.~Gambino and P.~A. Grassi, ``{The Nielsen identities of the SM and the
  definition of mass},''
  \href{http://dx.doi.org/10.1103/PhysRevD.62.076002}{{\em Phys. Rev.}
  {\bfseries D62} (2000) 076002},
\href{http://arxiv.org/abs/hep-ph/9907254}{{\ttfamily arXiv:hep-ph/9907254
  [hep-ph]}}.

\bibitem{Denner:1999gp}
A.~Denner, S.~Dittmaier, M.~Roth, and D.~Wackeroth, ``{Predictions for all
  processes e+ e- ---> 4 fermions + gamma},''
  \href{http://dx.doi.org/10.1016/S0550-3213(99)00437-X}{{\em Nucl. Phys.}
  {\bfseries B560} (1999) 33--65},
\href{http://arxiv.org/abs/hep-ph/9904472}{{\ttfamily arXiv:hep-ph/9904472
  [hep-ph]}}.

\bibitem{Denner:2006ic}
A.~Denner and S.~Dittmaier, ``{The Complex-mass scheme for perturbative
  calculations with unstable particles},''
  \href{http://dx.doi.org/10.1016/j.nuclphysbps.2006.09.025}{{\em Nucl. Phys.
  Proc. Suppl.} {\bfseries 160} (2006) 22--26},
\href{http://arxiv.org/abs/hep-ph/0605312}{{\ttfamily arXiv:hep-ph/0605312
  [hep-ph]}}.

\bibitem{Denner:2016etu}
A.~Denner, L.~Jenniches, J.-N. Lang, and C.~Sturm, ``{Gauge-independent
  $\overline{MS}$ renormalization in the 2HDM},''
  \href{http://dx.doi.org/10.1007/JHEP09(2016)115}{{\em JHEP} {\bfseries 09}
  (2016) 115},
\href{http://arxiv.org/abs/1607.07352}{{\ttfamily arXiv:1607.07352 [hep-ph]}}.

\bibitem{Altenkamp:2017ldc}
L.~Altenkamp, S.~Dittmaier, and H.~Rzehak, ``{Renormalization schemes for the
  Two-Higgs-Doublet Model and applications to h → WW/ZZ → 4 fermions},''
  \href{http://dx.doi.org/10.1007/JHEP09(2017)134}{{\em JHEP} {\bfseries 09}
  (2017) 134},
\href{http://arxiv.org/abs/1704.02645}{{\ttfamily arXiv:1704.02645 [hep-ph]}}.

\bibitem{Liebler:2010bi}
S.~Liebler and W.~Porod, ``{Electroweak corrections to Neutralino and Chargino
  decays into a W-boson in the (N)MSSM},''
  \href{http://dx.doi.org/10.1016/j.nuclphysb.2011.10.033,
  10.1016/j.nuclphysb.2011.03.018}{{\em Nucl. Phys.} {\bfseries B849} (2011)
  213--249}, \href{http://arxiv.org/abs/1011.6163}{{\ttfamily arXiv:1011.6163
  [hep-ph]}}.
[Erratum: Nucl. Phys.B856,125(2012)].

\bibitem{Cornwall:2010upa}
J.~M. Cornwall, J.~Papavassiliou, and D.~Binosi, {\em {The Pinch Technique and
  its Applications to Non-Abelian Gauge Theories}}.
\newblock Cambridge University Press, 2010.
\newblock
\url{http://www.cambridge.org/mw/academic/subjects/physics/particle-physics-and-nuclear-physics/pinch-technique-and-its-applications-non-abelian-gauge-theories?format=AR}.
\newblock

\bibitem{Staub:2008uz}
F.~Staub, ``{SARAH},'' {\em arXiv:0806.0538 [hep-ph]} (2008) ,
\href{http://arxiv.org/abs/0806.0538}{{\ttfamily arXiv:0806.0538 [hep-ph]}}.

\bibitem{Hahn:2000kx}
T.~Hahn, ``{Generating Feynman diagrams and amplitudes with FeynArts 3},''
  \href{http://dx.doi.org/10.1016/S0010-4655(01)00290-9}{{\em Comput. Phys.
  Commun.} {\bfseries 140} (2001) 418--431},
\href{http://arxiv.org/abs/hep-ph/0012260}{{\ttfamily arXiv:hep-ph/0012260
  [hep-ph]}}.

\bibitem{Hahn:1998yk}
T.~Hahn and M.~Perez-Victoria, ``{Automatized one loop calculations in
  four-dimensions and D-dimensions},''
  \href{http://dx.doi.org/10.1016/S0010-4655(98)00173-8}{{\em Comput. Phys.
  Commun.} {\bfseries 118} (1999) 153--165},
\href{http://arxiv.org/abs/hep-ph/9807565}{{\ttfamily arXiv:hep-ph/9807565
  [hep-ph]}}.

\bibitem{Dudenas:2017}
V.~Dūdėnas and T.~Gajdosik, ``{On the Renormalization of Neutrinos in the
  Seesaw Extension of the Two-Higgs Doublet Model},''
\href{http://dx.doi.org/10.5506/APhysPolB.48.2243}{{\em Acta Phys. Polon.}
  {\bfseries B48} (2017) 2243}.

\bibitem{Dreiner:2008tw}
H.~K. Dreiner, H.~E. Haber, and S.~P. Martin, ``{Two-component spinor
  techniques and Feynman rules for quantum field theory and supersymmetry},''
  \href{http://dx.doi.org/10.1016/j.physrep.2010.05.002}{{\em Phys.Rept.}
  {\bfseries 494} (2010) 1--196},
\href{http://arxiv.org/abs/0812.1594}{{\ttfamily arXiv:0812.1594 [hep-ph]}}.

\bibitem{Kniehl:2008cj}
B.~A. Kniehl and A.~Sirlin, ``{Pole Mass, Width, and Propagators of Unstable
  Fermions},'' \href{http://dx.doi.org/10.1103/PhysRevD.77.116012}{{\em Phys.
  Rev.} {\bfseries D77} (2008) 116012},
\href{http://arxiv.org/abs/0801.0669}{{\ttfamily arXiv:0801.0669 [hep-th]}}.

\bibitem{Kniehl:2014gfa}
B.~A. Kniehl, ``{Propagator mixing renormalization for Majorana fermions},''
  \href{http://dx.doi.org/10.1103/PhysRevD.89.116010}{{\em Phys. Rev.}
  {\bfseries D89} no.~11, (2014) 116010},
\href{http://arxiv.org/abs/1404.5908}{{\ttfamily arXiv:1404.5908 [hep-th]}}.

\bibitem{Espriu:2002xv}
D.~Espriu, J.~Manzano, and P.~Talavera, ``{Flavor mixing, gauge invariance and
  wave function renormalization},''
  \href{http://dx.doi.org/10.1103/PhysRevD.66.076002}{{\em Phys. Rev.}
  {\bfseries D66} (2002) 076002},
\href{http://arxiv.org/abs/hep-ph/0204085}{{\ttfamily arXiv:hep-ph/0204085
  [hep-ph]}}.

\bibitem{Davidson:2005cw}
S.~Davidson and H.~E. Haber, ``{Basis-independent methods for the
  two-Higgs-doublet model},''
  \href{http://dx.doi.org/10.1103/PhysRevD.72.099902,
  10.1103/PhysRevD.72.035004}{{\em Phys. Rev.} {\bfseries D72} (2005) 035004},
  \href{http://arxiv.org/abs/hep-ph/0504050}{{\ttfamily arXiv:hep-ph/0504050
  [hep-ph]}}.
[Erratum: Phys. Rev.D72,099902(2005)].

\bibitem{Haber:2006ue}
H.~E. Haber and D.~O'Neil, ``{Basis-independent methods for the
  two-Higgs-doublet model. II. The Significance of tan$\beta$},''
  \href{http://dx.doi.org/10.1103/PhysRevD.74.015018,
  10.1103/PhysRevD.74.059905}{{\em Phys. Rev.} {\bfseries D74} (2006) 015018},
  \href{http://arxiv.org/abs/hep-ph/0602242}{{\ttfamily arXiv:hep-ph/0602242
  [hep-ph]}}.
[Erratum: Phys. Rev.D74,no.5,059905(2006)].

\bibitem{Fleischer:1980ub}
J.~Fleischer and F.~Jegerlehner, ``{Radiative Corrections to Higgs Decays in
  the Extended Weinberg-Salam Model},''
\href{http://dx.doi.org/10.1103/PhysRevD.23.2001}{{\em Phys. Rev.} {\bfseries
  D23} (1981) 2001--2026}.

\bibitem{Actis:2006ra}
S.~Actis, A.~Ferroglia, M.~Passera, and G.~Passarino, ``{Two-Loop
  Renormalization in the Standard Model. Part I: Prolegomena},''
  \href{http://dx.doi.org/10.1016/j.nuclphysb.2007.04.021}{{\em Nucl. Phys.}
  {\bfseries B777} (2007) 1--34},
\href{http://arxiv.org/abs/hep-ph/0612122}{{\ttfamily arXiv:hep-ph/0612122
  [hep-ph]}}.

\bibitem{Krause:2016gkg}
M.~Krause, ``{On the Renormalization of the Two-Higgs-Doublet Model},''
  Master's thesis, KIT, Karlsruhe, TP, 2016.
\newblock
  \url{https://www.itp.kit.edu/_media/publications/masterthesismarcel.pdf}.

\bibitem{Krause:2016oke}
M.~Krause, R.~Lorenz, M.~Muhlleitner, R.~Santos, and H.~Ziesche,
  ``{Gauge-independent Renormalization of the 2-Higgs-Doublet Model},''
  \href{http://dx.doi.org/10.1007/JHEP09(2016)143}{{\em JHEP} {\bfseries 09}
  (2016) 143},
\href{http://arxiv.org/abs/1605.04853}{{\ttfamily arXiv:1605.04853 [hep-ph]}}.

\bibitem{Krause:2017mal}
M.~Krause, D.~Lopez-Val, M.~Muhlleitner, and R.~Santos, ``{Gauge-independent
  Renormalization of the N2HDM},''
  \href{http://dx.doi.org/10.1007/JHEP12(2017)077}{{\em JHEP} {\bfseries 12}
  (2017) 077},
\href{http://arxiv.org/abs/1708.01578}{{\ttfamily arXiv:1708.01578 [hep-ph]}}.

\bibitem{Denner:1991kt}
A.~Denner, ``{Techniques for calculation of electroweak radiative corrections
  at the one loop level and results for W physics at LEP-200},''
  \href{http://dx.doi.org/10.1002/prop.2190410402}{{\em Fortsch.Phys.}
  {\bfseries 41} (1993) 307--420},
\href{http://arxiv.org/abs/0709.1075}{{\ttfamily arXiv:0709.1075 [hep-ph]}}.

\end{thebibliography}\endgroup

\end{document}